\documentclass[sigconf]{acmart}
\AtBeginDocument{%
  }

\setcopyright{acmlicensed}
\copyrightyear{2018}
\acmYear{2018}
\acmDOI{XXXXXXX.XXXXXXX}
\acmConference[Conference acronym 'XX]{Make sure to enter the correct
  conference title from your rights confirmation email}{June 03--05,
  2018}{Woodstock, NY}
\acmISBN{978-1-4503-XXXX-X/2018/06}
\usepackage{multirow}
\usepackage{enumitem}
\usepackage{booktabs}

\begin{document}

\title{HyMiRec: A Hybrid Multi-interest Learning Framework for LLM-based Sequential Recommendation}

\author{Jingyi Zhou}
\affiliation{%
  \institution{Xiaohongshu Inc.}
  \city{Beijing}
  \country{China}
}
\affiliation{%
  \institution{Fudan University}
  \city{Shanghai}
  \country{China}
}

\author{Cheng Chen}
\affiliation{%
  \institution{Xiaohongshu Inc.}
  \city{Beijing}
  \country{China}
}

\author{Kai Zuo}
\affiliation{%
  \institution{Xiaohongshu Inc.}
  \city{Beijing}
  \country{China}
}

\author{Manjie Xu}
\affiliation{%
  \institution{Xiaohongshu Inc.}
  \city{Beijing}
  \country{China}
}
\affiliation{%
  \institution{Peking University}
  \city{Beijing}
  \country{China}
}

\author{Zhendong Fu}
\affiliation{%
  \institution{Xiaohongshu Inc.}
  \city{Beijing}
  \country{China}
}

\author{Yibo Chen}
\affiliation{%
  \institution{Xiaohongshu Inc.}
  \city{Beijing}
  \country{China}
}

\author{Xu Tang}
\affiliation{%
  \institution{Xiaohongshu Inc.}
  \city{Beijing}
  \country{China}
}

\author{Yao Hu}
\affiliation{%
  \institution{Xiaohongshu Inc.}
  \city{Beijing}
  \country{China}
}

\renewcommand{\shortauthors}{Zhou et al.}

\begin{abstract}
Large language models (LLMs) have recently demonstrated strong potential for sequential recommendation. However, current LLM-based approaches face critical limitations in modeling users' long-term and diverse interests. First, due to inference latency and feature fetching bandwidth constraints, existing methods typically truncate user behavior sequences to include only the most recent interactions, resulting in the loss of valuable long-range preference signals. Second, most current methods rely on next-item prediction with a single predicted embedding, overlooking the multifaceted nature of user interests and limiting recommendation diversity.
To address these challenges, we propose HyMiRec, a hybrid multi-interest sequential recommendation framework, which leverages a lightweight recommender to extracts coarse interest embeddings from long user sequences and an LLM-based recommender to captures refined interest embeddings. To alleviate the overhead of fetching features, we introduce a residual codebook based on cosine similarity, enabling efficient compression and reuse of user history embeddings. To model the diverse preferences of users, we design a disentangled multi-interest learning module, which leverages multiple interest queries to learn disentangles multiple interest signals adaptively, allowing the model to capture different facets of user intent.
Extensive experiments are conducted on both benchmark datasets and a collected industrial dataset, demonstrating our effectiveness over existing state-of-the-art methods. Furthermore, online A/B testing shows that HyMiRec brings consistent improvements in real-world recommendation systems. Code is available at https://github.com/FireRedTeam/FireRedSeqRec.
\end{abstract}




\maketitle

\section{Introduction}
Sequential recommendation (SR) aims to predict users’ next preferred interaction items based on their historical behavior sequence \cite{Wang2019SequentialRS,10.1145/3159652.3159656,10.1145/1772690.1772773}. As a classical user-to-item recommendation paradigm, SR has been widely adopted in various scenarios such as e-commerce, video streaming, online advertising, and personalized content feeds  \cite{Chang2021SequentialRW,zhang2025frequencyaugmented}. The accurate and holistic modeling of users’ latent interests has long been a core problem in sequential recommendation. Early SR methods primarily rely on ID-based item representations, using architectures such as recurrent neural networks (RNNs), attention layers, or lightweight Transformers to model user preferences ~\cite{Hidasi2015SessionbasedRW,Kang2018SelfAttentiveSR,10.1145/3638535,10.5555/3304222.3304315}. While these approaches have achieved notable success, they are often limited by their insufficient semantic understanding to unseen items and constrained modeling capacity over complex dependencies.

\begin{figure}[t]
\centering
\includegraphics[width=0.99\columnwidth]{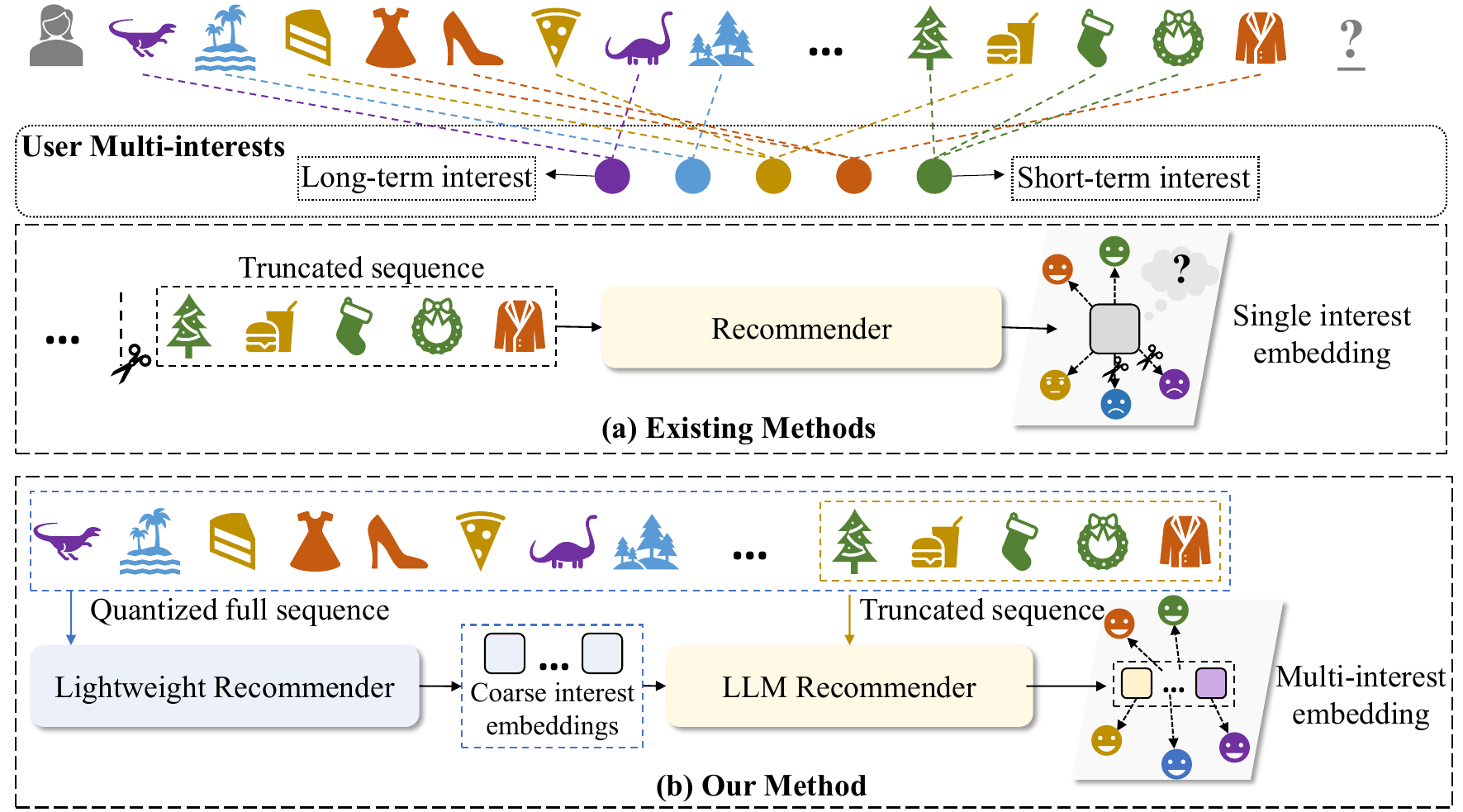} 
\caption{Comparison of our method and existing methods. A user often has multiple distinct interests, also including both long-term and short-term preferences. (a) Existing methods: truncate history sequences and use a single embedding to learn multiple interests, which may cause long-term interest forgetting and entangled representations. (b) Our method: adopts a hybrid framework to model long- and short-term behaviors with multi-interest embeddings, enabling effective learning of user's multiple interests.}
\label{fig1}
\end{figure}

Recently, large language models (LLMs) have shown great potential in sequential recommendation, demonstrating impressive performance owing to their exceptional contextual understanding capabilities and extensive world knowledge~\cite{Liu_Wu_Wang_Wang_Zhu_Zhao_Tian_Zheng_2025,liao2024llara,yin2025unleash}. Current efforts to leverage LLM for sequential recommendation can be broadly categorized into two paradigms: (1)Using LLMs as feature extractors to obtain rich content representations for items~\cite{Liu_Wu_Wang_Wang_Zhu_Zhao_Tian_Zheng_2025,10.1145/3539618.3591932}, though performance remains limited by downstream model capacity; (2) Employing LLMs as recommendation models to leverage their sequence modeling and reasoning abilities~\cite{liao2024llara}. Works like HLLM~\cite{HLLM} integrate both paradigms into an end-to-end trainable framework, yielding promising results. Despite these advances, existing methods still struggle to effectively modeling users’ long-term and diverse interests in complex real-world scenarios.

Two key challenges limit current approaches. First, high online queries per second (QPS) requirements create significant pressure on feature retrieval and real-time inference, forcing most methods to use truncated inputs. While this approach alleviates the computational burden, it inherently disregards long-term user behavior patterns, which are crucial for capturing evolving and persistent interests over time. Second, these methods struggle to learn diverse interests which are semantically different from single interest embedding, leading to homogenized recommendations biased toward recent interactions while neglecting niche or long-term interests. Moreover, the next-item training objective misaligns with real-world session-based recommendation, limiting the model’s ability to capture holistic user intent. Besides, in realistic recommendation scenarios, user clicks within a short window are often interchangeable, making next-item labels noisy and unstable.

To address the challenges, we propose HyMiRec, a hybrid framework for long-term multi-interest sequential recommendation leveraging LLMs. Our framework leverages a lightweight recommender to extracts coarse interest embeddings from long interaction histories to represent users’ long-term patterns and a LLM recommender to model refined user interest based on the combination of the coarse embedding and recent interactions. To reduce online retrieval costs, we further introduce a cosine-similarity-based residual codebook that compresses item embeddings into compact representations, leveraging their inherent clustering properties.
In order to comprehensively capture users’ diverse potential interests and enhance the diversity of retrieved items, we propose a Disentangled Multi-Interest Learning (DMIL) module. Instead of treating the next item as the only positive sample, DMIL considers a window of future interactions as positive signals, reflecting broader and more robust possible user intents. Items are clustered into multiple interest groups and matched with learnable interest queries, while a matching-based contrastive loss encourages the model to disentangle different user preferences effectively.

To validate the effectiveness of HyMiRec, we conduct extensive experiments on PixelRec8M, MovieLens 100K, and a large-scale industrial dataset with millions of items and interactions, demonstrating that our approach significantly outperforms existing methods across multiple metrics. Furthermore, online A/B testing in production environments confirms the practical impact of our framework, showing substantial improvements in recommendation personalization and diversity.

Our contribution can be summarized as follows:

\begin{itemize}
    \item We propose HyMiRec, a hybrid sequential recommendation framework that consists of a lightweight recommender and a LLM recommender to jointly model coarse long-term patterns and refined real-time interests.
    \item We introduce a disentangled multi-interest learning module to learn diverse user preferences with matching-based contrastive loss, where learnable queries are dynamically assigned to learn disentangled interests, enabling full and comprehensive learning. 
    \item Experimental results demonstrate that our method outperforms existing methods on public benchmarks and a industrial dataset. Online A/B results also confirm our improvements in real-world recommendation system.
\end{itemize}

\section{Related Works}
\subsection{LLM based Sequential Recommendation}
The application of large language models (LLMs) to sequential recommendation has recently attracted increasing attention due to their powerful understanding and generative ability. Existing works can be broadly categorized into two paradigms. The first type of work leverages LLMs as powerful feature extractors to enhance item representations~\cite{Liu_Wu_Wang_Wang_Zhu_Zhao_Tian_Zheng_2025,Yang2024SequentialRW,Liu2024LLMESRLL}. For example, LLMEmb~\cite{Liu_Wu_Wang_Wang_Zhu_Zhao_Tian_Zheng_2025} converts item attributes into text prompts, feeds them into an LLM to obtain the item representation. LLM-ESR~\cite{Liu2024LLMESRLL} further combines LLM-derived semantic embeddings with collaborative filtering-based embeddings to improve the expressive power of item representations. These methods benefit from the LLM's strong understanding of textual content and significantly improve the quality of learned item embeddings. However, their performance is still limited by the capability of the downstream recommender, which is typically lightweight and lacks sequence-level modeling capacity. 
The second category directly utilizes LLMs as sequential recommenders by formatting user behavior sequences into token-style inputs and generating the next item in a generative manner. A typical approach represents sequences using item ID embeddings, which are then fed into the LLM for next-item prediction~\cite{Bao2023TALLRecAE}. There are also works which concatenate the textual information (e.g., titles) of previous interacted items into a prompt and prompt the LLM to generate the next item description~\cite{Liao2025MultiGrainedPT}. In addition, HLLM ~\cite{HLLM} integrates both paradigms into an end-to-end trainable framework, jointly learning semantic item representations and sequential behaviors. Despite their promise, existing LLM-based approaches still face critical challenges in real-world scenarios—particularly in capturing users’ long-term and diverse interests, which limits their performance.

\subsection{User Multi-Interest Modeling}
Modeling users' diverse interests has been explored to improve recommendation diversity and accuracy. ComiRec~\cite{Cen2020ControllableMF} is one of the early attempts that introduces two approaches: (1) it uses multiple queries to attend to recent behaviors via self-attention, and (2) it employs capsule networks to extract interest vectors. Both approaches optimize for next-item prediction, where the target item is matched to the most relevant interest embedding for loss computation. In contrast, MIP~\cite{Shi2022EveryonesPC} introduced a cluster-aware attention mechanism, restricting attention computation to within item clusters to improve efficiency. The final interest representations were derived from the last positions of each cluster’s output sequence. Recent advances, such as Miracle~\cite{10.1145/3539618.3591778} and IMSR~\cite{10184671} use capsule networks with dynamic routing for interest separation, but are still constrained by the model scale. Kuaiformer~\cite{Liu2024KuaiFormerTR} further refined the multi-query approach with a Transformer-based framework, but similarly optimized for single-item prediction, where only the most relevant query received meaningful gradient updates. However, these methods typically rely on next-item prediction, which introduces noise and makes it hard to supervise diverse interests. Moreover, updating only the closest interest vector can lead to undertrained queries, which in turn harms overall performance.

\subsection{Long-term user interest modeling}
Modeling users' long-term sequences is recognized as beneficial for recommendation, which has been explored in several works. Memory-based methods~\cite{Ren2019LifelongSM,Pi2019PracticeOL,10.1145/3159652.3159668} use memory networks or memory banks to store and retrieve past interactions. Efficiency-oriented methods~\cite{Liu2023LinRecLA,Fan2024TiM4RecAE} aim to process long sequences more efficiently, such as reducing the computational complexity of the attention mechanism or model architecture. With the emergence of LLM-based sequential recommendation, handling long sequences poses new challenges. Real-time inference latency and feature retrieval bandwidth severely limit the feasibility of directly encoding full-length histories. While solutions are still limited, recent work has explored potential directions. ReLLa~\cite{Lin2023ReLLaRL} retrieves the top-k historical items most relevant to the target item, substantially reducing input length, but is limited to ranking tasks where the target item is known in advance. PatchRec~\cite{Liao2025MultiGrainedPT} divides the sequence into sessions and applies average pooling within each session to shorten the sequence. Yet, this coarse summarization introduces semantic loss and disrupts behavioral continuity. Therefore, effectively utilizing long sequences in LLM-based sequential recommendation remains a significant challenge in real-world scenarios.
\begin{figure*}[t]
\centering
\includegraphics[width=0.99\textwidth]{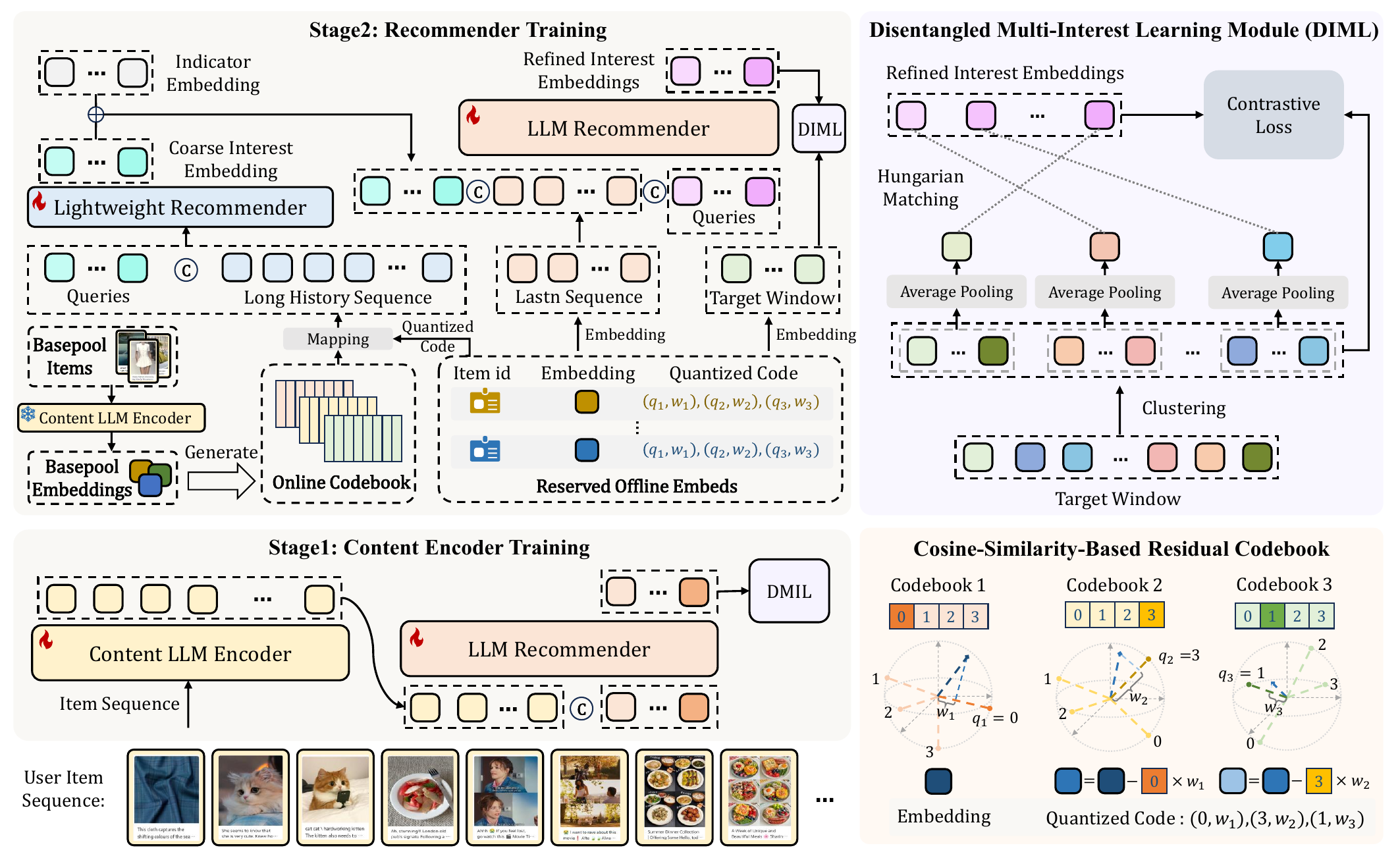} 
\caption{Overview of the proposed HyMiRec. In Stage 1, a content LLM encoder and an LLM recommender are jointly trained on short sequences, where the encoder extracts item embeddings and the recommender predicts user interests. The trained encoder is then fixed to build a cosine-similarity-based residual codebook and compress item embeddings into quantization codes. In Stage 2, long user histories are reconstructed from codes and processed by a lightweight recommender with learnable queries to yield coarse interests, which are combined with recent interactions and refined by the LLM recommender. The final interest embeddings are optimized with the proposed Disentangled Multi-Interest Learning (DMIL) module using future window targets.}
\label{fig2}
\end{figure*}

\section{Method}
\subsection{Overview}
Figure~\ref{fig2} illustrates the overview of our HyMiRec framework. The training process consists of two stages. In the first stage, we train a content LLM encoder based on a pre-trained LLM to extract item embeddings through end-to-end optimization with recommendation loss on recent interactions. The content LLM encoder extracts content embeddings for each item, while the LLM recommender takes the sequence of embeddings as input to predict the future interest embeddings of the user. After training, the content encoder is fixed to extract and store embeddings for all items offline. We then build a cosine-similarity-based residual quantization codebook using the embeddings of the base pool items randomly selected from all items. The embeddings of items are further compressed into quantization codes using the codebook for efficient storage and reuse. In the second stage, users' long historical sequences are reconstructed from quantized codes and the online codebook. The historical sequence combines with learnable queries as input to a lightweight recommender, producing coarse interest embeddings that capture long-term behavioral patterns. These coarse embeddings are then combined with an indicator embedding and concatenated with the recent sequence, forming the input to the LLM recommender to generate refined interest embeddings. Finally, the interest embeddings are optimized through our proposed Disentangled Multi-Interest Learning (DMIL) module using item embeddings from the target window.
\subsection{Hybrid Recommender Framework}
\subsubsection{Content LLM Encoder}
The content LLM encoder is used to extract rich semantic representations from each item, leveraging the LLM's powerful text comprehension capabilities. For each item, we concatenate its title and content text into a single input text. For example, the input text is formatted as:
"Title: When the Sun Meets the Horizon, content: The sky was so clear today, I couldn’t help but stop and watch the sunset. The colors were absolutely breathtaking, like nature's own masterpiece!"
After tokenizing this concatenated text, we append a special token at the end of the sequence to signal the importance of the resulting embedding. The sequence is then passed through the content LLM Encoder, where the output corresponding to the special token is extracted as the item embedding.
\subsubsection{Cosine-Similarity-based Residual Codebook for Efficient Representation}
To reduce inference resources and latency, we extract and store item embeddings offline in industrial scenarios. During online inference, the model retrieves historical interaction embeddings for user items. However, high QPS in production environments creates retrieval bandwidth bottlenecks that limit embedding fetching speed. To address this, we propose a cosine-similarity-based residual codebook that leverages the natural clustering and hierarchical structure in item embeddings. 

To construct the cosine-similarity-based residual codebook, we select a large set of items as basepool items and extract their item embeddings using the note LLM encoder, defined as $\{e_1,...,e_N\}$ where $N$ is the number of basepool items. The embeddings are then clustered via a multi-layer residual clustering process, which uses cosine similarity as the distance metric. Specifically, for the $i^{th}$ layer of the codebook, we perform balanced k-means clustering on the residual embeddings $\{e_1^i,...,e_N^i\}$ of the layer based on cosine similarity to obtain cluster centers $\{c_1^i,...,c_k^i\}$ as the centroids of the codebook $i$. 

We employ cosine similarity for clustering motivated by its alignment with industrial online retrieval systems where cosine distance is the standard metric for item recall, ensuring cluster assignments directly optimize for downstream recall accuracy. After the $i^{th}$ layer codebook is generated, for the $j^th$ embedding, the residual embeddings of the ${i+1}^{th}$ layer is calculated as:
\begin{equation}
    b_j^i=\mathop{\arg\max}\limits_{x \in [1,k]}\;cos(e_j^i,c_x^i)
\end{equation}
\begin{equation}
    e^{i+1}_j = e_j^i-\frac{e_j^i \cdot c_{b_j}^i}{||c_{b_j}^i||^2} \cdot  c_{b_j^i}^i , j \in[1,N]
\end{equation}
where $cos$ is the cosine similarity. Since cosine-distance-based codebook ignores the magnitude of vectors, we calculate projection residuals instead of direct residuals. This ensures that each layer operates within a subspace that is orthogonal to the previous layers' directions, as $e_j^{i+1} \perp c_{b_j}^i$, effectively reducing information redundancy. At the same time, the projection retains the magnitude information of the embedding along the direction of the cluster center, which enables more precise reconstruction during the follow-up process.
Finally, we obtain a 3-layer codebook $\mathbf{C}$ where a 2048-dimensional item embedding can be represented by three integers and three floating-point numbers, significantly reducing retrieval cost and memory footprint.
\subsubsection{Hybrid Recommender for Coarse-to-Fine Interests Modeling} In sequential recommendation, long-term user behavior sequences contain rich signals that reflect user preferences and behavioral patterns. However, these sequences are often highly repetitive and contain substantial amounts of irrelevant noise. Inputting the entire sequence directly into the LLM recommender will bring excessive memory overhead and computational cost. To balance the efficiency and granularity of user interest modeling, we propose a hybrid recommender that hierarchically processes user behavior sequences, reducing the computational burden while preserving essential behavioral signals.
For the long history interaction sequence $H^u=\{h_1^u,\ldots,h_l^u\}$ of a user $u$, we first retrieve the quantization codes $q_i$ and projections $p_i$ of the each item $h_i^u$ in sequence to reconstruct the long history sequence embedding $E^{ht}$. The sequence is concatenated with learnable queries $Q_{coarse}$ as the input of the lightweight recommender, a shallow Transformer, to obtain the coarse interest embeddings $R_{coarse}^u$ as follows:
\begin{equation}
    E^{ht} = [e^{ht}_1;\cdot\cdot\cdot;e^{ht}_l]
\end{equation}
\begin{equation}
    e^{ht}_i = \sum_{j=1}^{3} \mathbf{C}^j_{q_i} \cdot p_i
\end{equation}
\begin{equation}
    R_{coarse}^u=\phi_{Light}([E^{ht},Q_{coarse}])
\end{equation}
where $\phi_{Light}$ is the weight of the lightweight recommender and the $\mathbf{C}^j_{q_i}$ is the $q_i^{th}$ code in the $j^th$ layer of the codebook $\mathbf{C}$.

The coarse interest embeddings preaggregates the relevant information related to the current task, providing a compact representation that reflects the user's general interest tendencies. Then, we retrieve the embeddings of the last-n interactions to generate lastn embeddings $E^{ln}$, which provide valuable signals that reflect refined and immediate user intent. Besides, an indicator embedding $I$ is added to the coarse interest embeddings as a marker for the LLM to recognize them. Finally, the lastn embeddings combined with marked coarse interest embeddings and learnable queries $Q_{refined}$ are fed into the LLM recommender to generate refined interest embeddings $R_{refined}^u$. 
\begin{equation}
    E^{ln} = [e^{ln}_1;\cdot\cdot\cdot;e^{ln}_n]
\end{equation}
\begin{equation}
    R_{refined}^u=\phi_{LLM}([R_{coarse}+I,E^{ln},Q_{refined}])
\end{equation}
where $n$ is the length of last-n sequence and $\phi_{LLM}$ is the weight of the LLM recommender.
The lightweight recommender is jointly trained with the LLM recommender, allowing the coarse interest embeddings to learn to pre-aggregate task-relevant information from long sequences.

\subsection{Disentangled Multi-Interest Learning Module}
To effectively capture the diverse and potentially divergent interests of users, we propose Disentangled Multi-Interest Learning Module (DMIL). Instead of relying on a single interest embedding which often biased toward dominant interests, we encourage the model to attend to different facets of user intent.
\subsubsection{Window Targets Design}
Unlike conventional next-item prediction that treats a single next item as the positive sample, we consider all items within the subsequent window as positives, under the hypothesis that these items collectively reflect the user’s short-term interests without strict precedence. Besides, it better aligns with real-world inference scenarios, where users do not refresh the recommendation list after every single interaction. Instead, the goal is to surface a set of items that are likely to be of interest within the next viewing session, rather than predicting a strictly ordered next click.
\subsubsection{Disentangled Learning Loss}To model user interests in a disentangled manner, we introduce a set of learnable queries, each aimed at capturing a distinct refined interest. To ensure stable and effective learning of diverse interests, our design adheres to three principles: (1) each query should receive sufficient supervision, (2) queries should minimize overlap in interest coverage, and (3) individual queries should optimize toward coherent directions.

Concretely, given the refined interest embeddings \( \{\mathbf{r}_1,...,\mathbf{r}_s\} \) obtained by queries and positive samples \( \{\mathbf{t}_1,...,\mathbf{t}_w\} \) from the target window, we first cluster the window’s positive samples into \( s \) groups $\mathcal{G}=\{G_1, \ldots, G_s\}$ via cosine similarity, aligning the cluster count with the interests number. The \( s \) cluster centroids $\{\mathbf{g}_1,...,\mathbf{g}_s\}$ are then matched to \( s \) refined interest embeddings via Hungarian algorithm to maximize the pairwise cosine similarities:
\begin{equation}
    \max_{\Pi \in \mathcal{P}_s} \sum_{j=1}^s \cos(\mathbf{r}_j, \mathbf{g}_{\Pi(j)})
\end{equation}
where $\Pi$ is a permutation function.
Finally, the contrastive loss is employed only between each refined interest embedding and its matched positive samples:
\begin{equation}
    \mathcal{L}_{total}=\frac{1}{w}\sum_{i=1}^{w}\sum_{j=1}^{s} \mathcal{L}_{ctr}(\mathbf{t}_i,\mathbf{r}_j)\cdot \mathbb{I}[\mathbf{t}_i \in G_{\Pi(j)}]
\end{equation}
where $\mathbb{I}[\cdot]$ is an indicator function. $\mathcal{L}_{ctr}$ is the contrastive loss function, defined as:
\begin{equation}
\mathcal{L}_{ctr}(t, r) = -\log \frac{e^{cos(t, r)/\tau}}{e^{cos(t, r)/\tau}+\sum_{i=1}^{m} e^{{cos}(r, e_m)/\tau}}
\end{equation}
where $m$ is the number of negative samples randomly sampled , $e_i$ is the embedding of the $i^th$ negative sample, and $cos$ is the cosine similarity. Our DMIL module enables adaptive learning of different user’s interests. For users with multiple latent interests, different queries tend to specialize and learn distinct preference patterns. In contrast, for users with narrow or focused interests, the queries naturally converge to similar representations, avoiding unnecessary fragmentation.

\subsection{Inference and Online Serving Strategy}
During inference, we leverage the refined interest embeddings \( \{\mathbf{r}_1,...,\mathbf{r}_s\} \) to retrieve relevant items from the candidate pool \( \{\mathbf{a}_1,...,\mathbf{a}_{num}\} \). Specifically, each refined interest embedding independently computes cosine similarity scores with all candidate item embeddings. Assume that $cos_{ij}$ is the cosine similarity score between $a_i$ and $r_j$, the final relevance score $sim_i$ is determined as:
\begin{equation}
sim_i=\mathop{\arg\max}\limits_{j \in [1,s]}\;cos_{ij}
\end{equation}
The candidates are then sorted by their final scores, ensuring that an item matching any dominant interest receives a high recall probability.

For online serving, all item embeddings and their corresponding quantized codes are pre-computed offline using the trained content LLM encoder and the codebook, and then stored in the retrieval system. For each user, the refined interest embeddings are updated via online inference after every 10 new interactions, which balances personalization freshness with computational overhead.
To minimize the modification cost to existing retrieval infrastructure, we adopt a query-splitting strategy during candidate generation. Given a retrieval channel configured to return $K$ items, each interest embedding is used to independently retrieve $K$ candidates. The union of these candidates is then deduplicated, and final similarity scores are computed for all remaining candidates. The top $K$ items are selected as the final recommendation set.
\section{Experiments}
In this section, comprehensive experiments are conducted on public benchmarks and an industrial dataset, covering both offline and online settings, to evaluate the effectiveness of our proposed HyMiRec. Specifically, we aim to address the following research questions:
\begin{itemize}[left=0pt]
    \item \textbf{RQ1:} How does HyMiRec perform compared to existing state-of-the-art methods in both public benchmarks and industrial datasets?
    \item \textbf{RQ2:} Can HyMiRec consistently improve key online metrics in real-world recommendation systems?
    \item \textbf{RQ3:} How do the main components of HyMiRec contribute to its overall effectiveness?
    \item \textbf{RQ4:} How do different hyper-parameter settings affect the performance of HyMiRec?
\end{itemize}

\subsection{Experimental Settings}
\subsubsection{Dataset} We conduct offline evaluations on two public benchmark datasets,PixelRec~\cite{Cheng2023AnID} and MovieLens-1M~\footnote{https://movielens.org/}, as well as a real-world industrial dataset. To ensure data quality and enable a clearer comparison, we remain users with more than 60 interactions in both public datasets. For the industrial dataset, we randomly select 571,958 user click sequences from different user clusters and content topics. Detailed statistics of all datasets are summarized in Table ~\ref{dataset}.

\begin{table}
\centering
\begin{tabular}{c|cccc} 
\toprule
Dataset            & \#User & \#Item   & \#Avg. L. & Avg. T.  \\ 
\hline
PixelRec           & 148335 & 98833    & 51.38     & 64.39    \\
MovieLens          & 3938   & 3677     & 234.7     & 15.79    \\
Industrial & 571958 & 11708332 & 241.11    & 229.1    \\
\bottomrule
\end{tabular}
\caption{Statics of PixelRec, MovieLens-1M and our industrial dataset. L.: Length of the interaction sequence. T.: Token number  of the text.}
\label{dataset}
\end{table}

\subsubsection{Implementation Details} All experiments are conducted on Nvidia H800 GPUs. The code is implemented using Python 3.10.12 and PyTorch 2.3.0. We set the learning rate to 1e-5, and adopt the Adam optimizer along with a cosine learning rate scheduler with a warm-up ratio of 0.1. The batch size and the negative samples of each batch is set to 16 and 8 for public datasets and industrial dataset repectively. The number of refined interest embeddings is set to 2 for public datasets and 3 for industrial dataset, and the window size is 4 for public datasets and 8 for industrial dataset. The random seed is set to 2020. 
To ensure fair comparison, we uniformly adopt TinyLlama-1.1B~\cite{zhang2024tinyllama} as the backbone model for all LLM-based methods.

\begin{table*}
\centering
\begin{tabular}{c|c|cccc|cccc} 
\toprule
\multicolumn{2}{c|}{\multirow{2}{*}{Method}}  & \multicolumn{4}{c|}{PixelRec}                                         & \multicolumn{4}{c}{MovieLens-1M}                                       \\ 
\cmidrule{3-10}
\multicolumn{2}{c|}{}                         & R@10            & R@200           & N@10            & N@200           & R@10            & R@200           & N@10            & N@200            \\ 
\midrule
\multirow{3}{*}{ID-based methods~} & GRU4REC  & 0.0358          & 0.1646          & 0.02058         & 0.0429          & 0.2318          & 0.6846          & 0.1430          & 0.2197           \\
                                   & SASRec   & 0.0427          & 0.2137          & 0.0235          & 0.0532          & 0.2580          & 0.7016          & 0.1464          & 0.2304           \\
                                   & HSTU     & 0.0543          & 0.2422          & 0.0302          & 0.0631          & 0.2461          & 0.7296          & 0.1346          & 0.2263           \\ 
\midrule
\multirow{3}{*}{LLM-based methods} & HLLM     & 0.0583          & 0.2407          & 0.0329          & 0.0649          & 0.2715          & 0.6346          & 0.1562          & 0.2432           \\
                                   & Morec    & 0.0503          & 0.2241          & 0.0279          & 0.5824          & 0.2341          & 0.5863          & 0.1297          & 0.2161           \\
                                   & Patchrec & 0.0570          & 0.2417          & 0.0315          & 0.0639          & 0.2504          & 0.6302          & 0.1420          & 0.2328           \\ 
\midrule
\multicolumn{2}{c|}{HyMiRec(Ours)}            & \textbf{0.0608} & \textbf{0.2625} & \textbf{0.0337} & \textbf{0.0691} & \textbf{0.2811} & \textbf{0.7354} & \textbf{0.1607} & \textbf{0.2474}  \\
\bottomrule
\end{tabular}
\caption{The Comparision of our HyMiRec with existing SOTA methods on benchmark datasets, PixelRec and  MovieLens-1M. }
\label{public}
\end{table*}

\begin{table*}
\centering
\begin{tabular}{c|c|cccc|cccc} 
\toprule
\multicolumn{2}{c|}{Method}                   & R@10            & R@50            & R@100           & R@200           & N@10            & N@50            & N@100           & N@200            \\ 
\midrule
\multirow{3}{*}{ID-based methods}  & GRU4REC  & 0.0043          & 0.0197          & 0.0390          & 0.0664          & 0.0030          & 0.0055          & 0.0089          & 0.0118           \\
                                   & SASRec   & 0.0050          & 0.0213          & 0.0400          & 0.0690          & 0.0029          & 0.0052          & 0.0092          & 0.0120           \\
                                   & HSTU     & 0.0070          & 0.0237          & 0.0417          & 0.0747          & 0.0033          & 0.0068          & 0.0097          & 0.0133           \\ 
\midrule
\multirow{3}{*}{LLM-based methods} & HLLM     & 0.0163          & 0.0550          & 0.0827          & 0.1313          & 0.0085          & 0.0166          & 0.0210          & 0.0278           \\
                                   & Morec    & 0.0083          & 0.0267          & 0.0443          & 0.0774          & 0.0039          & 0.0078          & 0.0106          & 0.0152           \\
                                   & Patchrec & 0.0128          & 0.0477          & 0.0844          & 0.1347          & 0.0067          & 0.0141          & 0.0200          & 0.0271           \\ 
\midrule
\multicolumn{2}{c|}{HyMiRec(Ours)}            & \textbf{0.0227} & \textbf{0.0707} & \textbf{0.1047} & \textbf{0.1577} & \textbf{0.0115} & \textbf{0.0219} & \textbf{0.0274} & \textbf{0.0348}  \\
\bottomrule
\end{tabular}
\caption{The Comparision of our HyMiRec with existing SOTA methods on the Industrial dataset.}
\label{industrial}
\end{table*}

\subsubsection{Evaluation Metrics} To evaluate the recommendation performance, we adopt two widely-used metrics: Recall@K and NDCG@K. These metrics measure the relevance and ranking quality of the recommended items.
For PixelRec and MovieLens-1M, we use the full set of items as the candidate pool for evaluation. For the industrial dataset, the candidate pool consists of 414,480 items, which aligns with the realistic online serving environment.

\subsubsection{Baselines} To validate the effectiveness of our proposed method, we compare it with several competitive sequential recommendation baselines, including well-known ID-based methods, \emph{i.e.}, SASRec~\cite{Kang2018SelfAttentiveSR}, GRU4Rec~\cite{Hidasi2015SessionbasedRW}, and HSTU~\cite{Zhai2024ActionsSL}, as well as LLM-based methods, \emph{i.e.},  MoRec~\cite{10.1145/3539618.3591932}, HLLM~\cite{HLLM}, and PatchRec~\cite{Liao2025MultiGrainedPT}, where PatchRec is designed for long-term sequence modeling. The original PatchRec was designed for datasets where each item is represented only by a short textual title, failing to scale well in datasets where the number of items is large and each item has long textual content. Therefore, we re-implemented and adapted PatchRec to align with our setting, where the goal is to predict the content embedding of the next item rather than generating full text. To ensure a fair comparison, we constrain the input sequence length for LLM-based models to 10 on public benchmarks and 30 on industrial datasets, in consideration of inference latency. In contrast, non-LLM baselines are provided with longer sequences of 100 and 300 items, respectively.

\subsection{Offline Experiment (RQ1)}
The offline experiments demonstrate that HyMiRec consistently outperforms both ID-based and LLM-based methods across diverse datasets on almost all metrics. On public benchmarks MovieLens-1M and PixelRec, as shown in Table ~\ref{public}, HyMiRec achieves an average improvement of 24.53\% over the existing LLM-based methods on PixelRec and an average improvement of 8.9\% on MovieLens-1M, validating the overall effectiveness of HyMiRec. Notably, LLM-based methods generally surpass ID-based approaches attributed to their semantic understanding and sequential modeling capabilities.

The industrial dataset presents greater challenges, with a vast item space and high interaction sparsity, under which ID-based models suffer from limited generalization, as shown in Table \ref{industrial}. The performance gap becomes more pronounced in this setting, where our HyMiRec delivers an average improvement of 73.71\% over LLM baselines. This demonstrates the superior performance of our method in complex real-world environments, attributed to the coarse to refined hierarchical modeling of multiple user interests.

\begin{table*}
\centering
\begin{tabular}{c|cccc|cccc} 
\toprule
Method                      & R@10    & R@50   & R@100  & R@200  & N@10   & N@50    & N@100  & N@200   \\ 
\midrule
HyMiRec                     & 0.0227  & 0.0707 & 0.1047 & 0.1577 & 0.0115 & 0.0219  & 0.0274 & 0.0348  \\ 
\midrule
w/o lightweight recommender & 0.0207  & 0.064  & 0.1024 & 0.1494 & 0.0105 & 0.0199  & 0.0261 & 0.0326  \\
w/o Cosine-Similarity-based Residual Codebook                  & 0.0233  & 0.0714 & 0.1044 & 0.1580 & 0.0118 & 0.0221  & 0.0277 & 0.0350  \\ 
w/ Euclidean-Similarity-based Residual Codebook & 0.0213  & 0.0687  & 0.1027 & 0.1530 & 0.0108 & 0.0210  & 0.0267 &  0.0338 \\
w/o Indicator Embedding        &   0.0220  & 0.0694  & 0.1034 & 0.1547 & 0.0111 & 0.0216 & 0.0269  &  0.0342   \\
\midrule
w/o DIML                    & 0.0193  & 0.0624 & 0.0937 & 0.1474 & 0.0112 & 0.0208  & 0.0257 & 0.0333  \\
w/o window targets          & 0.0173  & 0.0597 & 0.0904 & 0.1427 & 0.0103 & 0.0202  & 0.0251 & 0.0312  \\
max mathcing                & 0.0180~ & 0.0610 & 0.0917 & 0.1450 & 0.0104 & 0.0200~ & 0.0255 & 0.0324  \\
\bottomrule
\end{tabular}
\caption{The ablation experimentes on the industrial dataset.}
\label{ablation}
\end{table*}

\subsection{Online A/B Experiment (RQ2)}
To further validate the effectiveness of our proposed HyMiRec framework in real-world environments, we conduct online A/B experiments in downstream recommendation scenarios, all situated in the recall stage of an industrial recommender system.

\subsubsection{Item Cold-start}
This scenario focuses on the cold-start of items published by users exposure in a two-column content feed, ultimately aiming to increase the publication intention of users. The effective delivery of newly published items will improve their visibility and interaction and finally incentivize the content publishers, creating a virtuous ecology. Accordingly, the key evaluation metrics include the daily publications and daily active publishers, which together reflect the system’s capability to stimulate content production and creator activity.

The experiment is conducted with a traffic allocation of 10\% vs. 10\% for at least one week. The recall candidate pool is restricted to items published within the past 24 hours with fewer than 1,000 impressions. As a result, HyMiRec yields a +0.44\% increase in daily publications and a +0.52\% increase in daily active publishers, indicating the enhanced creator activity.

\subsubsection{Ad Cold-start}
This scenario targets newly released advertising items, which often suffer from poor visibility and delayed engagement. Different from the first setting, the cold-start scenario for advertisements involves users who are advertisers with an explicit intention to publish. The core objective here is to maximize the reach and conversion potential of newly submitted ads. Thus, the core metric is cold-start pass-through rate—defined as the proportion of items that reach 500 impressions. A higher pass rate indicates a greater likelihood of exposure and thus a stronger potential for commercial conversion.
The experiment is conducted with a traffic allocation of 16\% vs. 16\% for at least one week. The recall pool includes ad items published within the past 3 days. We report results for two separate traffic source. In the image-and-text item traffic: pass-through rate improves from 26.46\% to 30.93\%. And in the short video item traffic, the pass-through rate increases from 13.19\% to 14.23\%. The results demonstrate HyMiRec’s superior ability to surface cold-start ads effectively, contributing to faster warm-up and improved commercial value.

\subsection{Ablation Study (RQ3)}
\subsubsection{Ablation for Hybrid Framework}
The ablation results of our hybrid architecture are shown in the upper part of Table.~\ref{ablation}. We remove the lightweight recommender and use only LLM-recommender with last-$n$ sequence to model user preferences, denote as w/o lightweight recommender. This leads to a consistent drop, indicating that incorporating the coarse interest embedding generated by the lightweight recommender provides a complementary signal to refined short-term modeling by LLM recommender. Moreover, to further validate the design of coarse interest embedding, we remove the indicator embedding that distinguishes different input sources, denoted as w/o Indicator Embedding. This variant results in a moderate decline, demonstrating that the indicator embedding indeed helps the LLM to differentiate input types. 

Furthermore, to evaluate the effectiveness of cosine-similarity-based residual codebook, we feed the original history sequence embedding directly into the lightweight recommender, skipping the quantization and reconstruction process of the cosine-similarity-based residual codebook, denote as w/o Cosine-Similarity-based Residual Codebook. While this variant yields slightly better performance, it introduce a significant system cost with the feature transmission bandwidth increases over 300×. Thus, our CSRC offers a favorable trade-off between performance and efficiency, enabling scalable deployment in real-world systems. 
Additionally, we replace the cosine-similarity-based codebook with an Euclidean-similarity-based codebook, denoted as w/ Euclidean-Similarity-based Residual Codebook, which leads to a noticeable accuracy drop.
As both the training loss and industrial retrieval rely on cosine similarity, cosine distance provides a more faithful measure of item similarity, whereas Euclidean distance leads to imprecise clustering assignments and loss of essential signals.  
Thus, our CSRC offers a favorable trade-off between performance and efficiency, enabling scalable deployment in real-world systems.

\subsubsection{Ablation for DMIL}
We further conduct ablation studies to assess the design of our Disentangled Multi-Interest Learning (DMIL) module, as shown in the lower part of Table~\ref{ablation}. The w/o DMIL setting removes the DMIL module and reverts to using a single learnable query to model user interest with the next item as the sole target. Performance drops significantly across all metrics, demonstrating the necessity of capturing diverse user intents through disentangled multi-interest learning. In w/o window targets, we retain the multi-query setup but use only the next item for supervision, where each target matches its most similar query for loss computation. This setting performs worse than the single-query variant due to supervision imbalance among multiple queries, degrading overall performance. In the max matching setting, we adopt the full multi-query and window target setup but replace our proposed disentangled learning with naïve max-matching: each window target is assigned to its most similar query. While this improves upon the single-target variant, it still suffers from training imbalance as some queries dominate the learning signal while others remain under-trained. In contrast, our DMIL module ensures sufficient and adaptive learning, leading to substantial performance gains.

\begin{figure}[t]
\centering
\includegraphics[width=0.99\columnwidth]{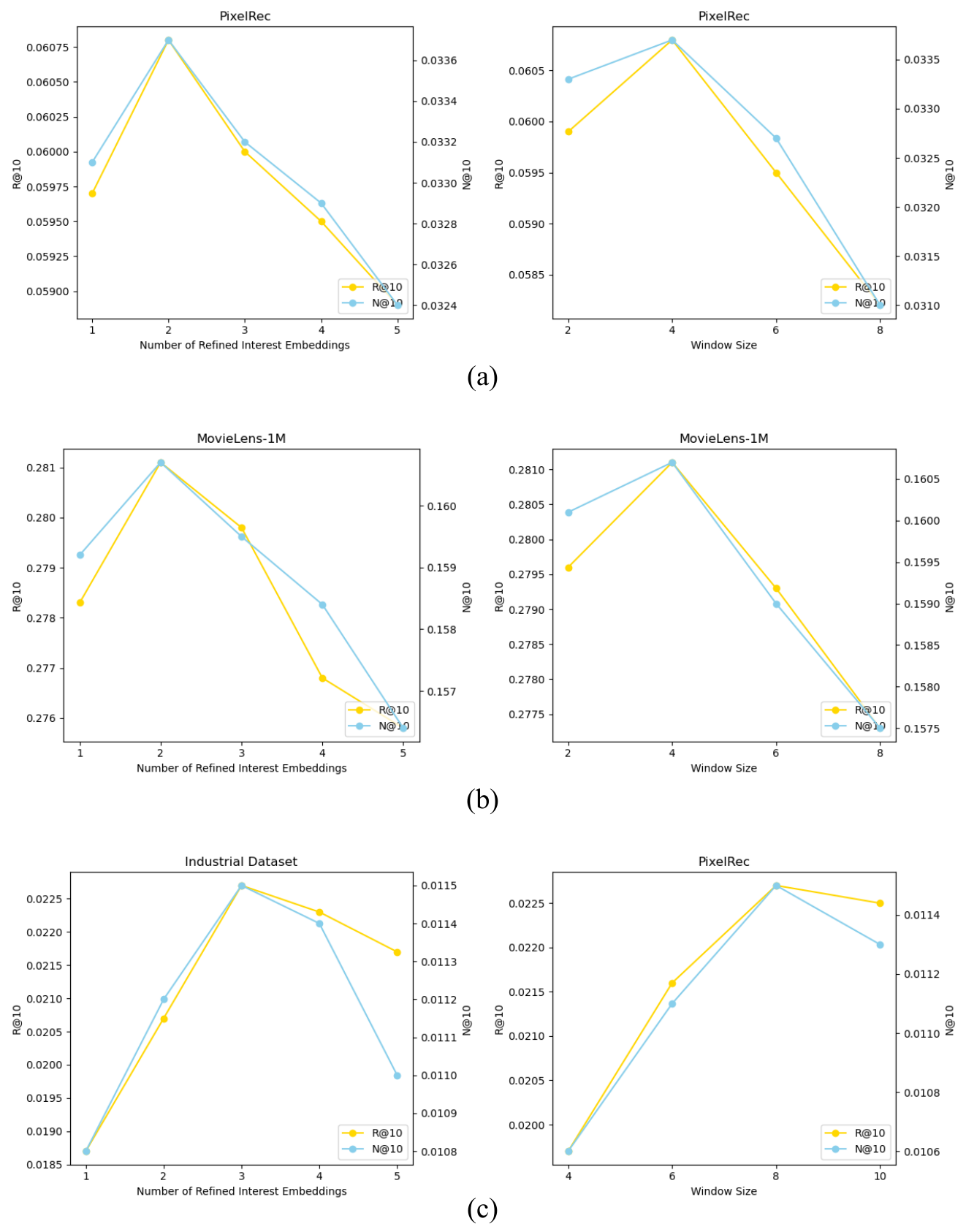} 
\caption{Performance of our HyMiRec under different hyper-parameters.}
\label{hyperparameter}
\end{figure}

\subsection{Hyper-Parameters Analysis (RQ4)}
To further investigate the impact of hyper-parameters to HyMiRec, we analyze two critical hyper-parameters on the three dataset: the number of refined interest embeddings and the window size for target sampling. 

The number of refined interest embeddings controls the capacity of the model to disentangle diverse user preferences. As shown in Figure~\ref{hyperparameter}, performance generally improves when increasing this number from small values, since more embeddings enable the model to capture richer facets of user intent. However, setting it too large leads to over-fragmentation of user preferences and unstable training, resulting in a performance drop. We also study the effect of window size, which determines how many future interactions are used as targets in the Disentangled Multi-Interest Learning module. A small window size limits the availability of positive samples and underestimates user preference continuity, while an excessively large window introduces potential noisy signals. 

Furthermore, it can be observed that the optimal number of refined interest embeddings and window size are larger in the industrial dataset compared to the public benchmarks. This is because our industrial dataset contains real user click logs where the signals are denser and the content is more diverse, resulting in the requirement of more refined embeddings to capture the heterogeneous and dynamic user preferences, and a larger window size to provide richer and more stable supervision signals. 

\section{Conclusion}
In this paper, we propose HyMiRec, a hybrid multi-interest sequential recommendation framework that addresses two critical challenges in modeling users' long-term and diverse interests. HyMiRec is a novel hybrid architecture that leverages a lightweight recommender for coarse long-sequence modeling and an LLM recommender for refined interest modeling, effectively balancing computational efficiency with modeling capacity. To reduce feature retrieval overhead, we introduce a cosine-similarity-based residual codebook that efficiently compresses historical interactions while preserving critical preference signals. We further propose a DMIL module that enables explicit modeling of multifaceted user interests through window-based targets and balanced multi-query learning. Extensive experiments on public benchmark datasets and a large-scale industrial dataset demonstrate that HyMiRec consistently outperforms strong baselines. Furthermore, online A/B testing across multiple real-world recommendation scenarios validates significant improvements in production environments.

\section{Future Work}
While HyMiRec demonstrates significant improvements in long-term and multi-interest sequential recommendation, several promising directions remain for future exploration. First, extending the residual codebook to support dynamic updates could further enhance its adaptability in real-world scenarios where user preferences evolve continuously. Second, investigating the integration of multimodal signals (e.g., text, images, audio) with sequential behavior data may unlock richer representations, particularly for content-rich platforms like short video and e-commerce. Additionally, we will further explore to unify HyMiRec with RL frameworks like PPO and DPO, optimizing long-term user satisfaction. 

\bibliographystyle{ACM-Reference-Format}
\bibliography{sample-base}

\appendix

\end{document}